\documentclass[intlimits,twoside,a4paper]{article}

\usepackage{xcolor}

\usepackage[cp1251]{inputenc}

\usepackage{cmpj3}
\usepackage{bm}

\issue{2021}{24}{2}{23801}
\doinumber{10.5488/CMP.24.23801}

\title[Structure and thermodynamics in the linear modified Poisson-Boltzmann equation]
{Structure and thermodynamics in the linear modified Poisson-Boltzmann theories in restricted primitive model electrolytes}
\author[L. B. Bhuiyan]{L. B. Bhuiyan\orcid{0000-0002-4574-173X}}

\address{ Laboratory of Theoretical Physics, Department of
Physics, University of Puerto Rico, 17 Avenida Universidad, STE 1701, San Juan, Puerto
Rico 00925-2537, USA
}


\date{Received January 4, 2021, in final form February 3, 2021}
\begin{document}

\maketitle
\begin{abstract}
Structure and thermodynamics in restricted primitive model
electrolytes are examined using three recently developed versions of a
linear form of the modified Poisson-Boltzmann equation. Analytical expressions
for the osmotic coefficient and the electrical part of the mean activity
coefficient are obtained and the results for the osmotic and the mean activity
coefficients are compared with that from the more established mean spherical
approximation, symmetric Poisson-Boltzmann, modified Poisson-Boltzmann theories,
and available Monte Carlo simulation results. The linear theories predict the
thermodynamics to a remarkable degree of accuracy relative to the simulations
and are consistent with the mean spherical approximation and modified
Poisson-Boltzmann results. The predicted structure in the form of the radial
distribution functions and the mean electrostatic potential also compare well
with the corresponding results from the formal theories. The excess internal
energy and the electrical part of the mean activity coefficient are shown to
be identical analytically for the mean spherical approximation and the linear
modified Poisson-Boltzmann theories.

\printkeywords
\end{abstract}

\section{Introduction}\label{s1}

    One of the more enduring theories in the physics and chemistry of Coulomb fluids
over the past (nearly) hundred years has been the theory of Debye and H\"{u}ckel (DH)~\cite {debyehuckel}, which is the linearized form of the classical Poisson-Boltzmann (PB)
theory. The intuitive simplicity of the DH concept together with the ease
of its implementation have been the theory's main attractions. For instance, almost all
variables required for a structural and thermodynamic description of an electrolyte solution
occur in closed forms in the DH and the Debye-H\"{u}ckel Limiting Law (DHLL) theory~\cite{mcquarrie}. Formal statistical mechanical analysis (see for example,~\cite{kirkwood})
and subsequent machine simulations~\cite{card,rasaiah,valleau1,valleau2,rogde,abramo}
over the years have brought out the deficiencies of the DH, the principal ones being the neglect
of the ionic exclusion volume and the ionic correlation terms. Some of the more
recent, salient references, and reviews are given by~\cite{vlachy,levin,henderson,messina,kalyuzhnyi,chersty}.

    The potential approach to the theory with its origins in the DH mechanism has evolved
over the decades through the pioneering work of Kirkwood  in the 1930s~\cite{kirkwood} and later
through the works of other authors~\cite{outh1,outh2,outh8,martinez,molero,outh5,outh6}
to the modified Poisson-Boltzmann (MPB) equations of today (see for example, reference~\cite{outh6}).
A popular alternate route is based on the liquid structure integral
equations such as the hypernetted chain (HNC)~\cite{friedman1,friedman2} and the mean spherical
approximation (MSA)~\cite{blum1,outh7,blum2}. The density functional theory (DFT) has also been
explored~\cite{attard,hansenlowen}.

    A widely used physical model used in conjunction with the above statistical mechanical theories in
studies of electrolytes has been the primitive model (PM), viz., arbitrary sized charged rigid spheres
moving in a dielectric continuum~\cite{vlachy}. The solvent is thus structureless being characterized
by a dielectric constant or relative permittivity $\varepsilon _{r}$. If the ion sizes are equal, then
we have the restricted primitive model (RPM). The RPM is also the underlying
model of the DH theory, since a RPM with the vanishing ion radius, except perhaps for the size of the
central ion, would lead to the latter. The HNC and the MPB
have been two of the most successful theories of PM or RPM electrolytes in the electrolyte solution regime
having been applied to a wide variety of situations under different physical conditions. The MPB formalism set
in the PM (or the RPM) yields a highly non-linear differential equation whose solution requires
involved numerical techniques~\cite{outh8} and thus may not be readily available. Fortunately, as
Outhwaite~\cite{outh9} showed a linear version of the MPB equation is tractable analytically leading
to a closed form expression for the mean electrostatic potential $\psi $.

Since the analysis of Outhwaite stated above, little has been reported in the literature on linear
theories based on the MPB, although a lot of work has been done with the MSA, viz., the works by Blum
(see for example, references~\cite {blum1,blum2}), and by Outhwaite and Hutson~\cite{outh7}, apart
from the obvious DH. In addition to being easier to use, linear theories
can offer valuable insights and understanding of the properties of
systems being examined albeit at the cost of a little accuracy. Linear solutions can also be valuable
in iterative numerical solution of corresponding non-linear equations. In a recent paper,
Outhwaite and Bhuiyan~\cite{outh10} studied the linear MPB (LMPB)
in some detail and formulated three versions of the equation, LMPB$i$, the index $\emph{i}$ ($\emph{i}=$ 1,2,3)
referring to special characteristics of a particular equation. Significantly, these equations
yielded an analytical solution for the~$\psi$, which, for 1:1 valency RPM electrolytes, compared
well with that from the MSA and MPB for a range of concentrations, and symmetric Poisson-Boltzmann (SPB)
theory~\cite{outh4} at low concentrations. Linearization retained the aspects of the MPB fluctuation
potential terms since the linear $\psi $'s also showed damped oscillations at higher concentrations
beyond the critical $y_{c}(=\kappa a)=$ 1.2412 ($\kappa $ being the Debye-H\"{u}ckel constant and
$a$ the common ionic diameter). In simple terms, fluctuation potential is the potential formulation
of the inter-ionic correlations and such oscillations are signatures for inter-ionic correlations,
which are not seen with the mean-field DH or the SPB.

In view of the comparative behaviour of LMPB $\psi $ with that from the SPB,
MPB, and MSA theories for a broad range of concentrations of RPM electrolytes
seen in~\cite{outh10}, we thought it of interest to apply the LMPB approach to an analysis of the
structure and thermodynamics in these systems. As we will see later, the LMPB expressions for
thermodynamic quantities such as the osmotic coefficient $\phi $ and the electrical contribution
to the mean activity coefficient $\gamma _{\pm}^{(\text{el})}$  develop into closed analytical forms,
which make their numerical evaluation straightforward. Addition of the hard-core component
$\gamma _{\pm}^{(\text{HS})}$ to the $\gamma _{\pm}^{(\text{el})}$ leads to the (full) $\gamma _{\pm}$.
It is worth mentioning here that the knowledge of $\phi $ and $\gamma _{\pm}$ has practical
significance in many chemical processes involving electrolyte solutions in industry and bio-sciences
(see for example,~\cite{barthelkrienkekunz, collinsneilsonenderby}). Relevant to this work,
we note also that recently Qui\~{n}ones et al. \cite{quinones} made extensive comparisons
of the SPB and MPB $\phi $ and $\gamma _{\pm}$ with the corresponding RPM or PM Monte Carlo~(MC)
data of Abbas~et~al.~\cite{abbas1,abbas2} for a wide range of solution concentrations.
The MPB, and to a lesser extent, the SPB showed a very good agreement with the simulations.
It would be interesting to see how well the LMPB predictions compare with these results.
Being experimentally measurable~\cite{robinson}, the measured $\phi $ and $\gamma _{\pm}$
can also provide a good assessment of theories.

    The organization of this paper is as follows. In the following section we
outline the principal equations of the LMPB theories pertinent to the calculation of
structure and thermodynamics of electrolytes. In section~\ref{s3} we present and discuss
the results of this work, while in section~\ref{s4} some general conclusions are drawn.

\section{Model and methods}

\subsection{Model}

    The model electrolyte system employed in this work is an aqueous RPM
electrolyte at around room temperature. This is consistent with one of the models
used by Abbas et al.~\cite{abbas1,abbas2} in their MC simulations, the other being
the PM.

    The various particle-particle interaction potentials are
\begin{equation}
 u_{ij}(r)=\left\{
\begin{array}{cc}
 \infty & r<(r_{i}+r_{j}) \\
\frac{e^{2} Z_i Z_j}{(4\piup \varepsilon _{0}\varepsilon _{r}r)} & r>(r_{i}+r_{j})
\end{array}
\right..
\label{m2_1}
\end{equation}
Here, $Z_{s}$ and $r_{s}$ are the valency and radius of ion species $s$, while
$r$ is the separation between two ions of types $i$ and $j$, respectively. Parameter
$\varepsilon _{0}$ is the vacuum permittivity and $|e|$ is the magnitude of the electronic charge.
For the RPM we have $r_{s} = a/$2 for all $s$.

\subsection{Methods}
\begin{figure}[!t]
	\centerline{\includegraphics[height=4.5in]{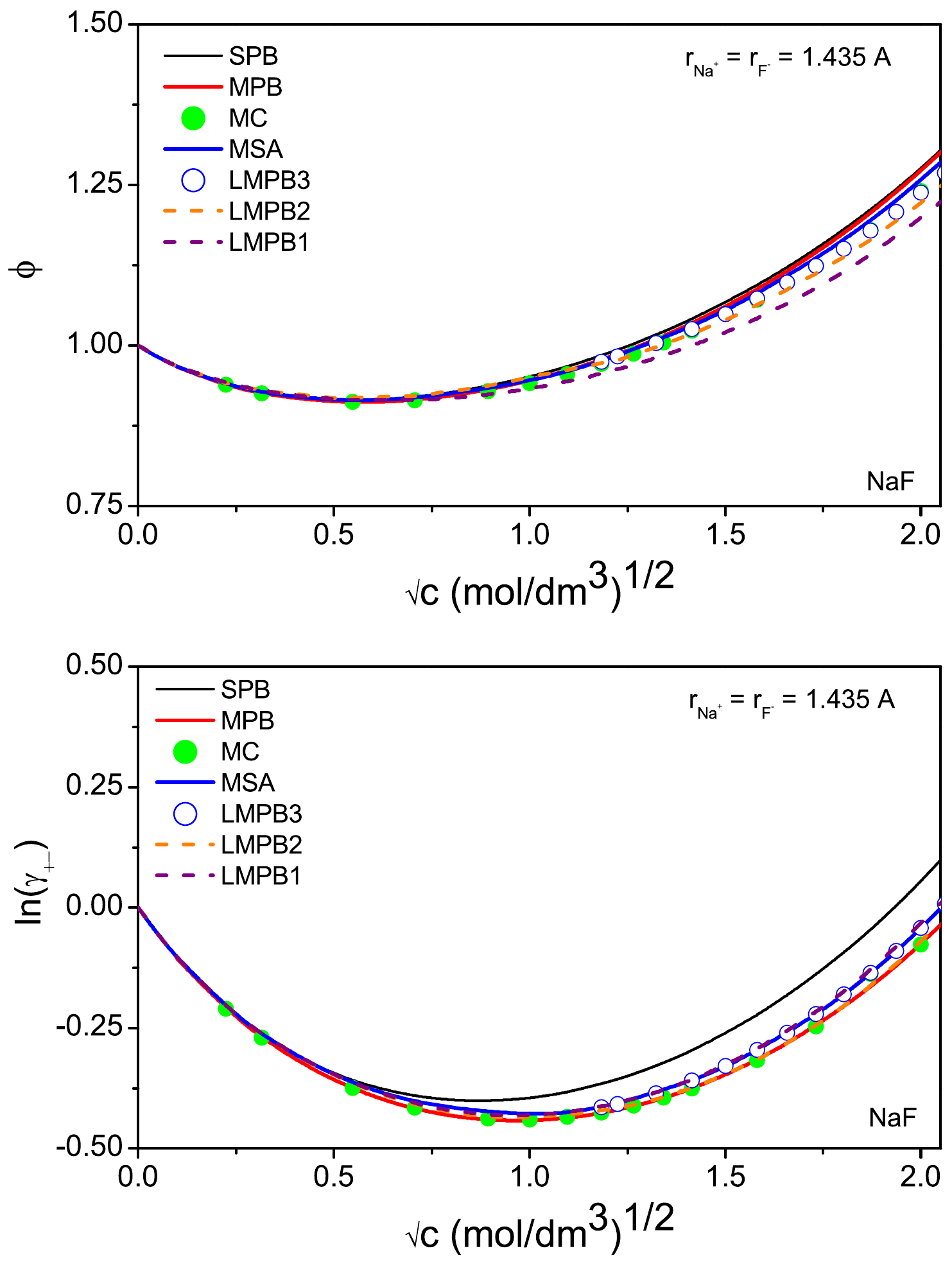}}
	\caption{(Colour online) MC and theoretical --- LMPB1, LMPB2, LMPB3, MSA, SPB, and MPB
		osmotic coefficient (upper panel), and natural logarithm of the mean activity coefficients (lower panel)
		as functions of the square root of the solution concentration for NaF using RPM. In the theoretical
		calculations, the common ionic radius is $r_{\text{Na}^{+}} = r_{\text{F}^{-}} =$~1.435~$\times 10^{-10}$~m, taken from
		the MC simulation data of references~\cite{abbas1,abbas2}. The MC data are from the same references.}
	\label{fig1}
\end{figure}
   
   The linearization of the MPB equation and the subsequent development of the
LMPB equations for the RPM electrolyte have been discussed in details in~\cite{outh10} and
will not be repeated here. We restrict ourselves to outlining the salient
equations in these theories.

    The Poisson equation for the $\psi _{i}(1;2)$ at the field point {\bf r}$_{2}$
in presence of an ion $i$ at {\bf r}$_{1}$ is

\begin{equation}
\nabla ^{2}\psi _{i}(1;2)=-\frac{|e|}{\varepsilon _{0}\varepsilon_{r}}\sum _{s}Z_{s}\rho _{s}g_{is}(1,2).
\label{m2_2}
\end{equation}

\noindent Here, $g_{is}(1,2)$ is the radial distribution function for the ion pair
$i$ and $s$ separated by $r_{is}( = r) =|{\bf r}_{1}-{\bf r}_{2}|$, with $\rho _{s}$ being
the mean number density of  ions of type $s$. The operator $\nabla $ operates on the
coordinates of the field point 2. It is convenient to use the transformation
$u_{i} = r \psi _{i}(1;2)$, whence the above equation transforms to

\begin{equation}
\frac{\rd^{2}u_{i}}{\rd r^{2}} = -\frac{|e|}{\varepsilon _{0}\varepsilon _{r}}\sum _{s}Z_{s}\rho _{s}rg_{is}(r).
\label{m2_3}
\end{equation}

    In the MPB approximation, the $g_{is}$ has been developed for the RPM as (see for example,
reference~\cite{outh6})

\begin{equation}
g_{is} = g_{is}^{0} \exp\left\{-\left(\frac{|e|\beta}{2}\right)\left[ Z_{s} L(u_{i})+Z_{i} L(u_{s})\right]\right\}.                                                          
\label{m2_4}
\end{equation}

\noindent The quantity $g_{is}^{0}$ above is the exclusion volume term, which is
the radial distribution function for two discharged ions in a sea of fully charged
ions, viz., $g_{is}^{0} = g_{is} (Z_{i} = Z_{s} = 0)$, $\beta = 1/(k_\mathrm{B}T)$, with $k_{\mathrm{B}}$
the Boltzmann's constant and $T$ the temperature. The operator $L(u)$ is given by

\begin{equation}
L(u)= \frac{1}{2r(1+y)}\left[u(r+a)+u(r-a)+ \kappa \int_{r-a}^{r+a}u(R)\rd R\right],
\label{m2_5}
\end{equation}

\noindent where $y = \kappa a$, and
$\kappa = [(e^{2}\beta/(\varepsilon_{0}\varepsilon_{r}))\sum_{s} Z_{s}^{2} \rho_{s}]^{1/2}$.

    Linearization of the non-linear MPB equation [equation~(\ref{m2_4}) substituted in~(\ref{m2_3})] above yields the
following

\begin{equation}
\frac{\rd^{2}u_{i}(r)}{\rd r^{2}} = g_{ij}^{0}\kappa ^{2} r L(u),    \hspace{0.5cm}      r \geqslant a.                                                                                    
\label{m2_6}
\end{equation}

 Within the charge free space $0 < r < a$, the solution of the
Laplace equation $\frac{\rd^{2}u_{i}(r)}{\rd r^{2}}= 0$ is as follows:

\begin{equation}
u_{i}(r) = r\left(\frac{\rd u_{i}}{\rd r}\right)_{r = a}+\frac{|e|Z_{i}}{4\piup \varepsilon_{0}\varepsilon_{r}},    \hspace{0.5cm}  0 \leqslant r \leqslant a.
\label{m2_7}
\end{equation}

\noindent Equation~(\ref{m2_6}) is valid for both symmetric and asymmetric valency systems. For instance,
for the linear theory with equal ion sizes we have $Z_{i}u_{j} = Z_{j}u_{i}$, so that in using equation~(\ref{m2_4})
in equation~(\ref{m2_3}) with the expression~(\ref{m2_5}) for $L(u)$, we have the terms such as $Z_{i}L(u_{j})$,
which can be written as $Z_{j}L(u_{i})$. Hence, the equation follows.

    The general solution of the linear MPB equation is governed by the roots of
a transcendental equation (cf. reference~\cite{outh10})

\begin{equation}
z \cosh(z) + y \sinh(z)= z^3 \frac{(1+y)}{y^{2}}.
\label{m2_8}
\end{equation}

 Taking the first two roots with the smallest real part, as these give the
physical solution, we can see that for small $y$ there are two real roots
that coalesce at the critical $y_{c} =$~1.2412 for symmetric valencies.
For $y > y_{c}$ the roots form a complex conjugate pair becoming imaginary at
a second critical point $y_{I} =$~7.83.

    Apart from the continuity of $u$ and $\rd u/\rd r$ at boundaries such as $r = a$,
there exists the exact condition, viz., the local electroneutrality

\begin{equation}
Z_{i}=-4\piup \sum _{s}Z_{s}\rho _{s}\int _{a}^{\infty} r^{2}g_{is}(r)\rd r.
\label{m2_9}
\end{equation}

\noindent Another useful relation is the Stillinger-Lovett (SL) second moment condition \cite{stillinger},
although it does not hold near the critical point of the electrolyte. The SL condition
can be written as \cite{outh11}

\begin{equation}
|e|\beta \sum_{s}Z_{s} \rho _{s}\int \psi _{s}\rd V = 1.                                              \label{m2_10}                                     \end{equation}

\noindent In~\cite{outh10}, the solutions to the LMPB equation were classified depending on
the boundary conditions and the exact conditions a solution satisfies.

\subsubsection{LMPB1, LMPB2 and LMPB3 equations}
\begin{figure}[!t]
		\centerline{\includegraphics[height=4.5in]{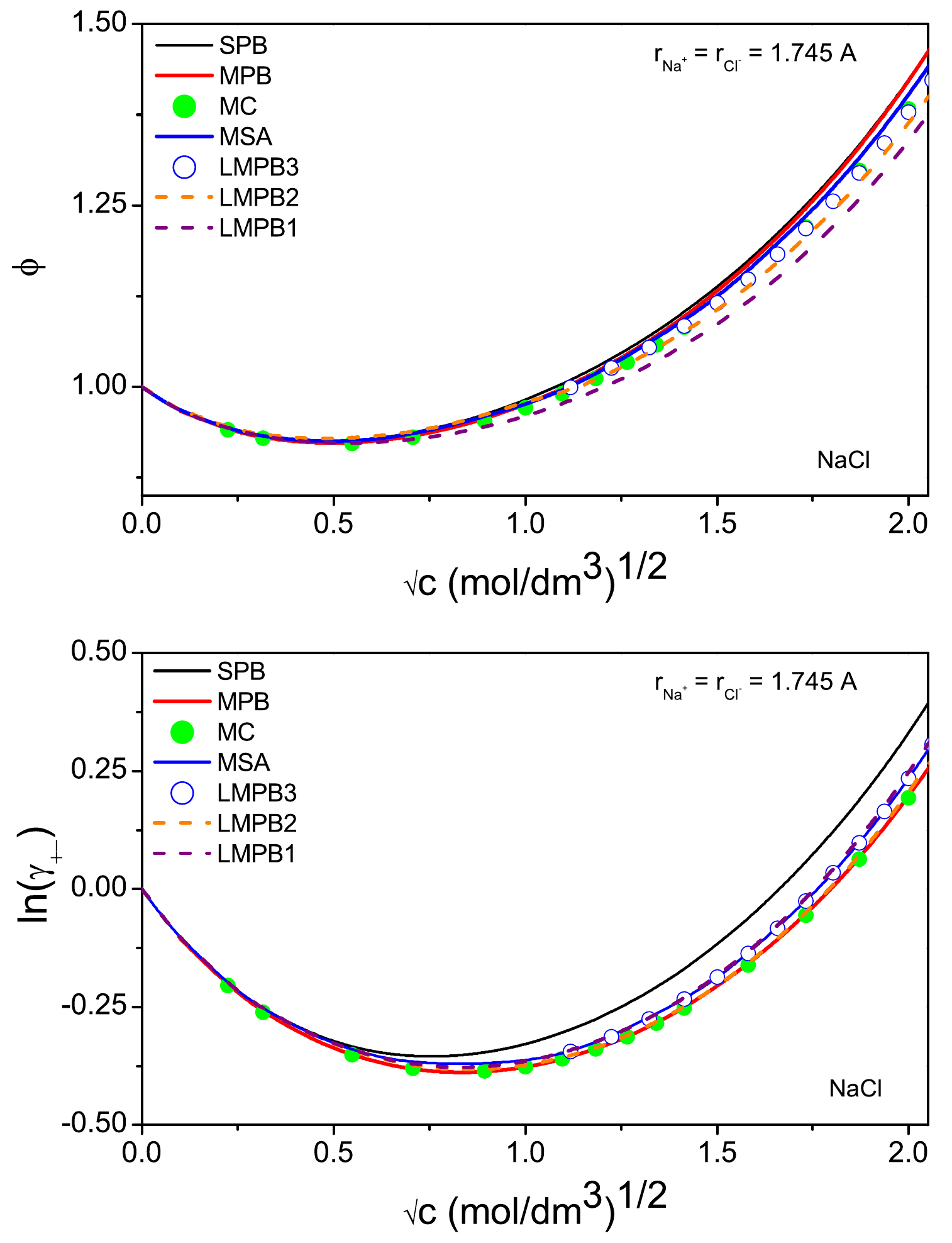}}
	\caption{(Colour online) MC and theoretical --- LMPB1, LMPB2, LMPB3, MSA, SPB, and MPB
		osmotic coefficient (upper panel), and natural logarithm of the mean activity coefficients (lower panel)
		as functions of the square root of the solution concentration for NaCl using RPM. In the theoretical
		calculations, the common ionic radius is $r_{\text{Na}^{+}} = r_{\text{Cl}^{-}} =$~1.745~$\times 10^{-10}$~m, taken from
		the MC simulation data of references~\cite{abbas1,abbas2}. The MC data are from the same references.}
	\label{fig2}
\end{figure}

\begin{figure}[!t]
	\centerline{\includegraphics [height=4.5in]{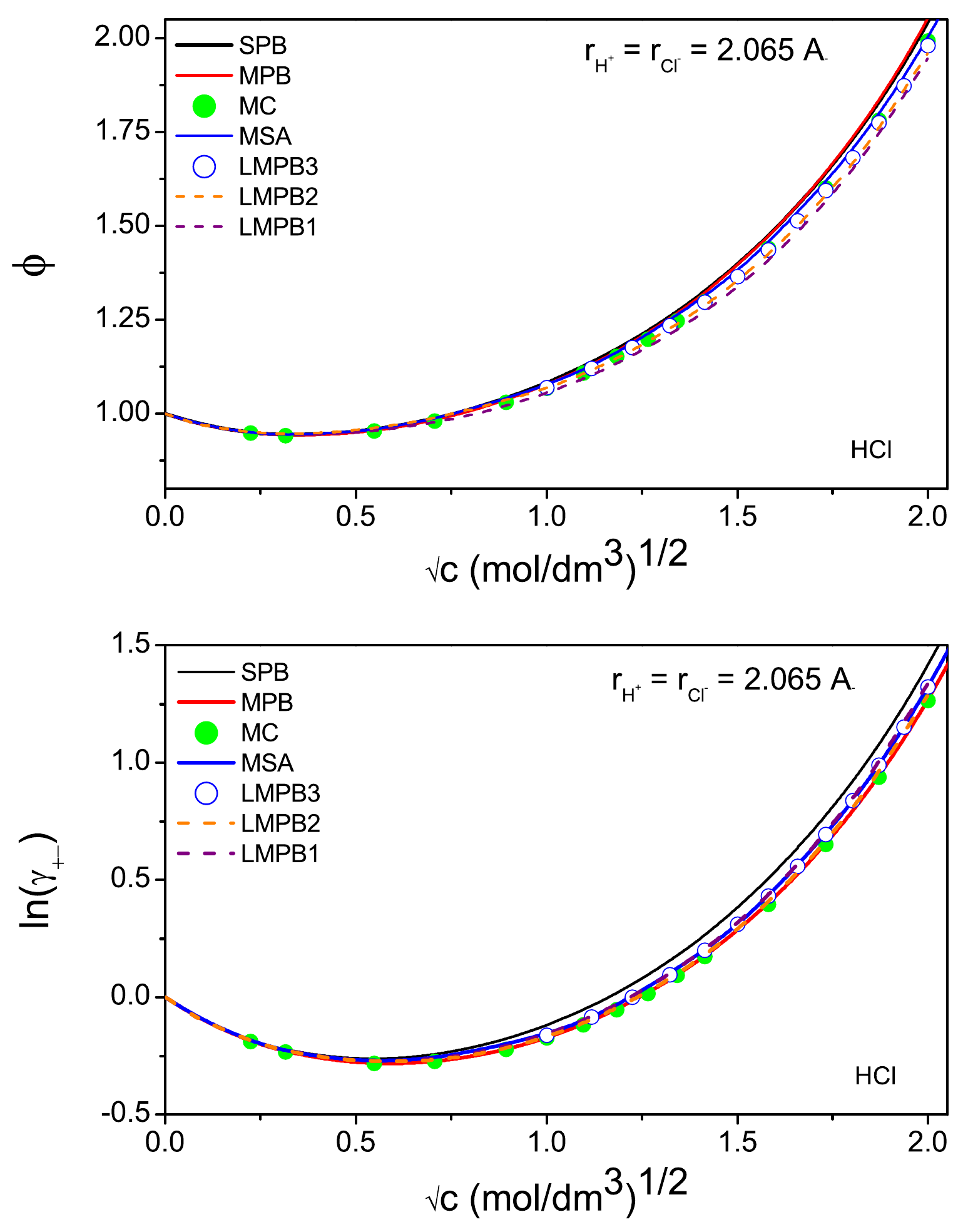}}
	\caption{(Colour online) MC and theoretical --- LMPB1, LMPB2, LMPB3, MSA, SPB, and MPB
		osmotic coefficient (upper panel), and natural logarithm of the mean activity coefficients (lower panel)
		as functions of the square root of the solution concentration for HCl using RPM. In the theoretical
		calculations, the common ionic radius is $r_{\text{H}^{+}} = r_{\text{Cl}^{-}} =$~2.065~$\times 10^{-10}$~m, taken from
		the MC simulation data of references~\cite{abbas1,abbas2}. The MC data are from the same references.}
	\label{fig3}
\end{figure}
    The LMPB1 equation satisfies the electroneutrality and the SL conditions,
while the LMPB2 satisfies the neutrality and the continuity of $u(r)$ at $r = a$.
In addition to the electroneutrality, and the SL conditions, the LMPB3 satisfies
the continuity of $u(r)$ and $\rd u/\rd r$ at both $r = a$ and $r = 2a$. This occurs
since a more accurate solution for $u(r)$ can be derived in the region
$a \leqslant r \leqslant  $ 2$a$, for example, by using the linear solution~(\ref{m2_7}) in equation~(\ref{m2_5})
to obtain $L$ in $a \leqslant r \leqslant  $ 2$a$. We refer the reader to~\cite{outh10} for further details.
\begin{figure}[!t]
	\centerline{\includegraphics [height=4.5in]{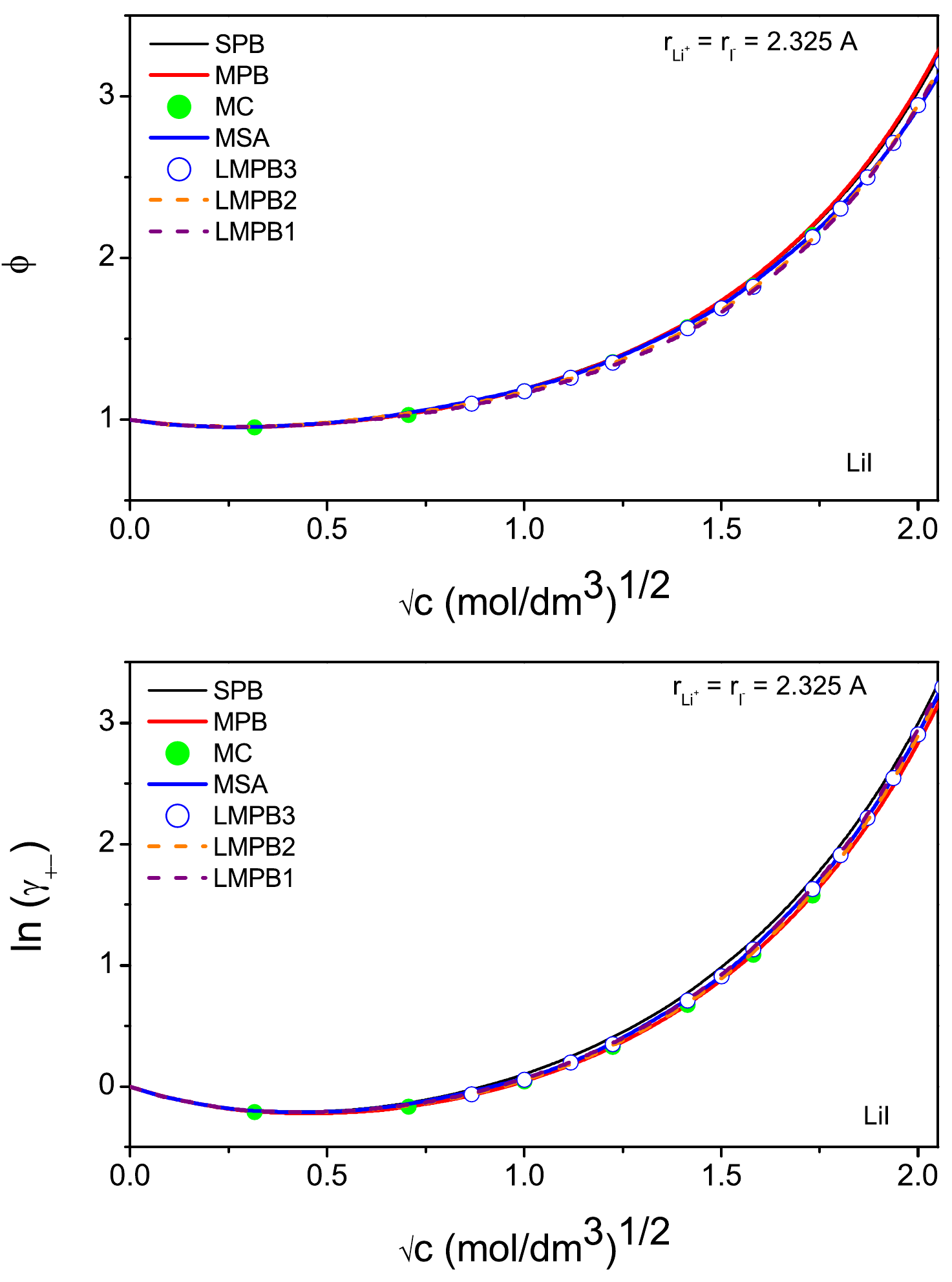}}
	\caption{(Colour online) MC and theoretical --- LMPB1, LMPB2, LMPB3, MSA, SPB, and MPB
		osmotic coefficient (upper panel), and natural logarithm of the mean activity coefficients (lower panel)
		as functions of the square root of the solution concentration for LiI using RPM. In the theoretical
		calculations, the common ionic radius is $r_{\text{Li}^{+}} = r_{\text{I}^{-}} =$~2.325~$\times 10^{-10}$~m, taken from
		the MC simulation data of references~\cite{abbas1,abbas2}. The MC data are from the same references.}
	\label{fig4}
\end{figure}

For $ y \leqslant y_{c}$, the LMPB1 and LMPB2 solutions can be written as

\begin{equation}
u_{i} = \frac{Z_{i}|e|}{4\piup \varepsilon _{0}\varepsilon _{r}}\left[A_{1}\exp(-\alpha _{1}r/a)+A_{2}\exp(-\alpha _{2}r/a)\right],
        \hspace{0.5cm}   r \geqslant  a,
\label{m2_11}
\end{equation}

\noindent where $\alpha _{1}$, $\alpha _{2}$ are the two real roots of the transcendental
equation~(\ref{m2_8}). The constants A$_{1}$, A$_{2}$ take on different forms for LMPB1 and LMPB2.
For instance, for LPMB1 we have

\begin{eqnarray}
&&A_{1}  =  \exp(\alpha _{1})[\alpha _{1}^{2}G_{2}-\omega (1+\alpha _{2})]/D, \nonumber\\
&&A_{2}  =  \exp(\alpha _{2})[-\alpha _{2}^{2}G_{1}+\omega (1+\alpha _{1})]/D, \nonumber\\&&
D  = \alpha _{1}^{2}(1+\alpha _{1})G_{2}-\alpha _{2}^{2}(1+\alpha _{2})G_{1}, \hspace{0.5cm} \omega = 6\alpha _{1}^{2}\alpha _{2}^{2}/y^{2}, \nonumber\\
&&G_{j}  =  \alpha _{j}^{3}+3\alpha _{j}^{2}+6\alpha _{j}+6, \hspace{0.5cm}  j = 1,2,
\label{m2_12}
\end{eqnarray}
 while for LMPB2 we have
\begin{eqnarray}
&&A_{1}  =  b_{2}\exp(\alpha _{1})/D, \nonumber\\
&&A_{2}  =  -b_{1}\exp(\alpha _{2})/D, \nonumber\\
&&D  =  b_{2}(1+\alpha _{1}) - b_{1}(1+\alpha _{2}), \nonumber\\
&&b_{j}  =  \frac{2\alpha _{j}^{2}}{\lambda }-(2+y)\alpha _{j}-2(1+y)-\frac{2y}{\alpha_{j}}-2\left(1-\frac{y}{\alpha _{j}}\right)\exp(-\alpha _{j}),  \hspace{0.5cm} j = 1,2, \nonumber\\
&&\lambda  =  \frac{y^{2}}{2(1+y)}. 
\label{m2_13}
\end{eqnarray}

\noindent The LMPB3 solution for this range of $y$ has not been given in~\cite{outh10}.

    For the range $y_{c} < y < y_{I}$, again the LMPB1 and LMPB2 solutions have the common form

\begin{equation}
u_{i} = (|e|Z_{i}/4\piup \varepsilon _{0}\varepsilon _{r})A\exp[-\alpha (r/a-1)]\cos[\beta(r/a-1)-B],  \hspace{0.5cm}   r \geqslant a,
\label{m2_14}
\end{equation}

\noindent where $\alpha $ and $\beta $ are now the real and imaginary parts of the complex
conjugate pair of roots of equation~(\ref{m2_8}). Furthermore, $A = \sqrt{X^{2}+Y^{2}}/D$, $B = \tan ^{-1}(Y/X)$,
$R = \alpha ^{2} - \beta ^{2}$ and $S = \alpha ^{2} + \beta ^{2}$.

    In the LMPB1 the constants $X$, $Y$ are

\begin{eqnarray}
&&G  = G_{1} + \ri G_{2},\nonumber\\
&&G_{1}  = \alpha ^{3} -3\alpha \beta ^{2}+3\alpha ^{2}-3\beta ^{2} +6\alpha +6, \nonumber\\
&&G_{2}  = 3\alpha ^{2}\beta -\beta ^{3} +6\alpha \beta +6\beta, \nonumber\\
&&H  = \left(\frac{6}{y^{2}}\right)S^{2}, \nonumber\\
&&X  = \beta H-RG_{2}+2\alpha \beta G_{1}, \nonumber\\
&&Y  = (1+\alpha )H-RG_{1}-2\alpha \beta G_{2}, \nonumber\\
&&D  = (1+\alpha )X-Y\beta.
\label{m2_15}
\end{eqnarray}

    For the LMPB2 we have from equation~(\ref{m2_12})  $b_{1}=Y+\ri X,  b_{2}=Y-\ri X$, so that

\allowdisplaybreaks\begin{eqnarray}
Y & = &\frac{2R}{\lambda}-(2+y)\alpha -2(1+y)-\frac{2y\alpha }{S}-2\exp(-\alpha )(p\cos\beta +q\sin\beta), \nonumber\\
X & = &\frac{4\alpha \beta }{\lambda }-(2+y)\beta +2q-2\exp(-\alpha )(q\cos\beta  -p\sin\beta  ), \nonumber\\
D & = &(1+\alpha )X-Y\beta, \nonumber\\
\lambda  & = &\frac{y^{2}}{2(1+y)}, \hspace{0.25cm} p=1-\frac{y\alpha }{S}, \hspace{0.25cm} q=\frac{qy}{S}.
\label{m2_16}
\end{eqnarray}

For the LMPB3 and for the range $y_{c} < y < y_{I}$ , the solution is

\begin{eqnarray}
u_{i} & = & \left(\frac{|e|Z_{i}}{4\piup \varepsilon _{0}\varepsilon _{r}}\right)\left(\frac{y^{2}}{2(1+y)}\right)\left[DX+EY+\mu (r)+c_{8}\eta (r)\right], \hspace{0.25cm} a \leqslant r  \leqslant 2a, \nonumber\\
u_{i} & = & \left(\frac{|e|Z_{i}}{4\piup \varepsilon _{0}\varepsilon _{r}}\right) A \exp(-\alpha r/a)\cos{(\beta r/a -B)}, \hspace{0.25cm} r \geqslant 2a,
\label{m2_17}
\end{eqnarray}

\noindent where $A = 2\sqrt{(X^{2}+Y^{2})}$ and $B = \tan ^{-1}(Y/X)$.
The quantities $X$, $Y$, $D$, $E$, $\mu (r)$ and $\eta (r)$ are very involved and we refer the
reader to~\cite{outh10} for details.

\subsubsection{Structure and thermodynamics}
    For the LMPB formulations, the pair distribution $g_{ij}$ can be constructed as

\begin{equation}
g_{ij}(r) = g_{ij}^{0}\exp\left\{-\frac{|e|Z_{j}}{g_{ij}^{0}\kappa ^{2}r}\left(\frac{\rd^{2}u_{i}}{\rd r^{2}}\right)\right\},
\label{m2_18}
\end{equation}

\noindent which gives a non-linear $g_{ij}$.
Another possibility would be to use a linearized version of the above, viz.,

\begin{equation}
g_{ij}^{\text{linear}}(r) = g_{ij}^{0}-\frac{|e|Z_{j}}{\kappa ^{2}r}\left(\frac{\rd^{2}u_{i}}{\rd r^{2}}\right).
\label{m2_19}
\end{equation}
The expression~(\ref{m2_18}) is analogous to the
DHX and EXP theories~\cite{mcquarrie,outh8,carley,card,outh12}, while the expression~(\ref{m2_19}) is
analogous to the MSA~\cite{outh7}. To avoid confusion, in the rest of this paper we refer to this linear form as LMPBi~(linear) $g_{ij}$ ($i =$~1,~2,~3).

    A consistency check on the $g$'s may be carried out through the integral

\begin{equation}
\bar{u_{i}} = \frac{|e|}{\varepsilon _{0}\varepsilon _{r}}\sum _{s}Z_{s}\rho _{s}\int_{r}^{\infty}(r-t)tg_{is}(t)\rd t.
\label{m2_20}
\end{equation}

We note that differentiating this twice with respect to $r$ immediately yields the requisite
Poisson's equation for $\bar{u_{i}}$. Hence, the extent of agreement between $u_{i}$ and $\bar{u_{i}}$
for an LMPB theory would be a measure of the consistency of the particular $g_{ij}$.

    The osmotic coefficient $\phi $ can be calculated from the relation (see for example, reference~\cite{outh8})

\begin{equation}
\phi = 1 + \frac{2\piup }{3\rho }\sum_{i}\sum_{s}\rho _{i}\rho _{s}g_{is}(a)a^{3} + \frac{\beta U^{(\text{ex})}}{3\rho},
\label{m2_21}
\end{equation}

\noindent where $\rho = \sum_{i}\rho _{i}$ and $U^{(\text{ex})}$ is the excess internal energy, viz.,

\begin{equation}
U^{(\text{ex})} = \frac{|e|}{2}\sum_{s}Z_{s}\rho _{s}\left (\psi _{s}(a)-\frac{|e|Z_{s}}{4\piup \varepsilon _{0}\varepsilon _{r}a}\right).
\label{m2_22}
\end{equation}

    Calculation of the activity coefficient is most conveniently achieved through the G\"{u}nteberg
charging process where the ion at the origin is charged up from zero to its full charge in a sea of
charged ions. The individual ionic activity has been derived as~\cite{outh8,molero}

\begin{equation}
\ln \gamma _{i} = \ln \gamma _{i}^{(\text{HS})} + \ln \gamma _{i}^{(\text{el})}.
\label{m2_23}
\end{equation}

\noindent The hard sphere part $\ln \gamma _{s}^{(\text{HS})}$ was analyzed by
Ebeling and Scherwinski~\cite{ebelsche}, while the charging process gives for the
electrical part~\cite{molero}

\begin{equation}
\ln \gamma _{i}^{(el)} = |e|Z_{i}\beta \int_{0}^{1}\lim _{r\rightarrow 0}\left(\psi _{i}-\frac{\lambda |e|Z_{i}}{4\piup \varepsilon _{0}\varepsilon _ {r}r}\right)\rd \lambda.
\label{m2_24}
\end{equation}
\begin{figure}[!t]
	\centerline{\includegraphics [height=4.5in]{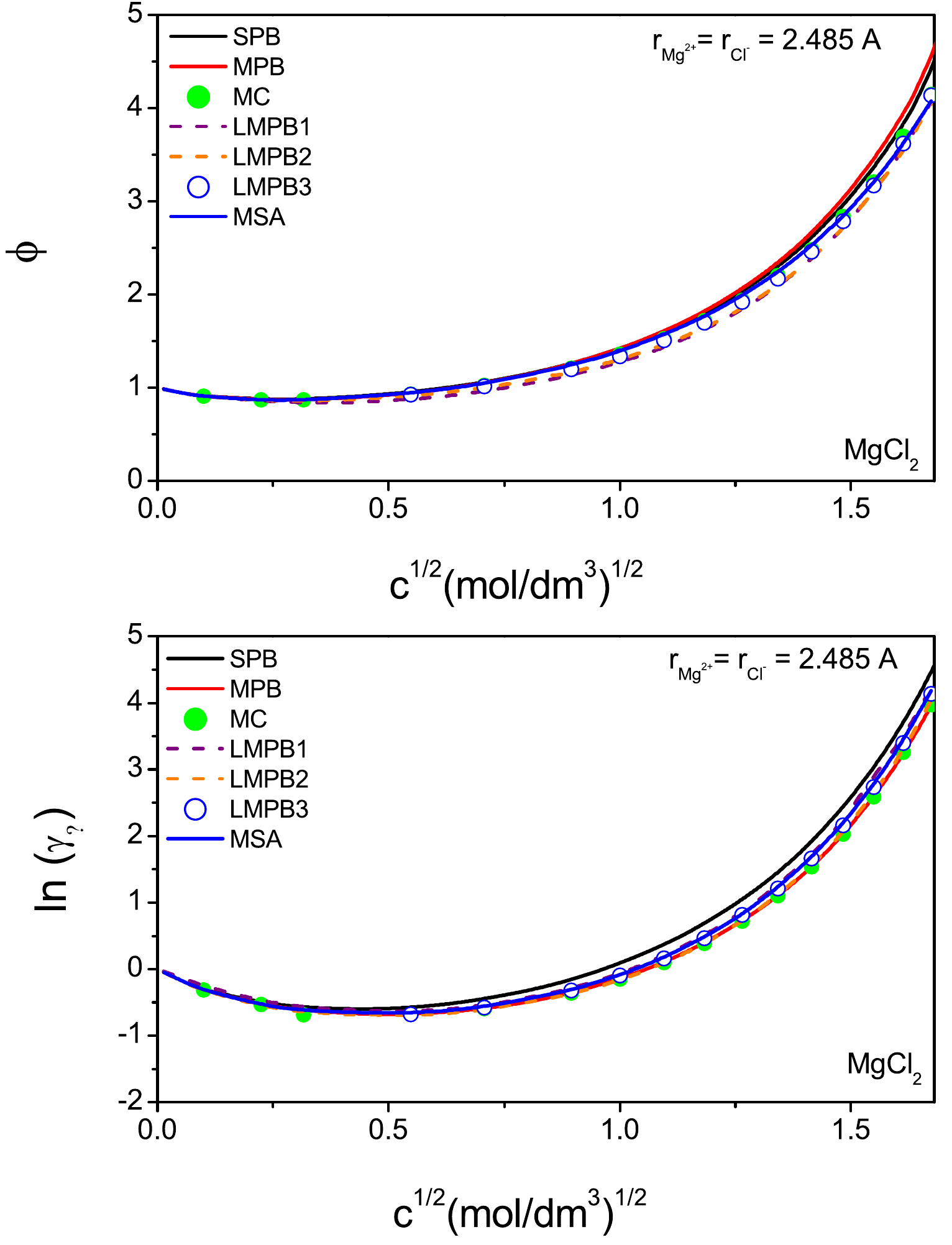}}
	\caption{(Colour online) MC and theoretical --- LMPB1, LMPB2, LMPB3, MSA, SPB, and MPB
		osmotic coefficient (upper panel), and natural logarithm of the mean activity coefficients (lower panel)
		as functions of the square root of the solution concentration for MgCl$_{2}$ using RPM. In the theoretical
		calculations, the common ionic radius is $r_{\text{Mg}^{2+}} = r_{\text{Cl}^{-}} =$~2.485~$\times 10^{-10}$~m, taken from
		the MC simulation data of references~\cite{abbas1,abbas2}. The MC data are from the same references.}
	\label{fig5}
\end{figure}

    Using now equations~(\ref{m2_7}), (\ref{m2_24}) with the LMPB solutions~(\ref{m2_11}), (\ref{m2_14}) and~(\ref{m2_17}) it is
straightforward to calculate the electrical part of the individual activity coefficient
for these theories. For example, for the LMPB1 and LMPB2 we have

\begin{equation}
 \ln \gamma _{i}^{(\text{el})} = \left\{
\begin{array}{cc}
-\frac{1}{2}Z_{i}^{2}\Gamma \left(A_{1}\alpha _{1}\exp (-\alpha _{1})+A_{2}\alpha _{2}\exp (-\alpha _{2})\right)  & y < y_{c} \\
-\frac{1}{2}Z_{i}^{2}\Gamma \left(A(\alpha \cos B - \beta \sin B)\right)   &   y_{c} < y < y_{I}
\end{array}
\right.,
\label{m2_25}
\end{equation}

\noindent where $\Gamma = (|e|^{2}\beta )/(4\piup \varepsilon _{0}\varepsilon _{r} a)$, with the
rest of the constants being given in equations~(\ref{m2_12}),~(\ref{m2_15}) for LMPB1 and equations~(\ref{m2_13}),~(\ref{m2_16})
for LMPB2.

    For LMPB3 we have

\begin{equation}
\ln \gamma _{i}^{(\text{el})} = \frac{1}{2}Z_{i}^{2}\Gamma \left(\frac{y^{2}}{2(1+y)}\right)\left\{X\left(\frac{\rd D}{\rd x}\right)_{x=1}+Y\left(\frac{\rd E}{\rd x}\right)_{x=1}+\left(\frac{\rd \mu }{\rd x}\right)_{x=1}+c_{8}\left(\frac{\rd \eta }{\rd x}\right)_{x=1}\right\},
\label{m2_26}
\end{equation}

\noindent with $x = r/a$. The mean activity coefficient follows from the individual coefficients~\cite{mcquarrie}

\begin{equation}
\ln \gamma _{\pm } = \frac{|Z_{+}|\ln \gamma _{-} + |Z_{-}|\ln \gamma _{+}}{|Z_{+}| + |Z_{-}|} .
\label{m2_27}
\end{equation}

\subsubsection{Equivalence of ln$\gamma _{\pm}^{(\text{el})} = \beta U^{(\text{ex})}/\rho$}
\begin{figure}[!t]
	\centerline{\includegraphics [height=4.5in]{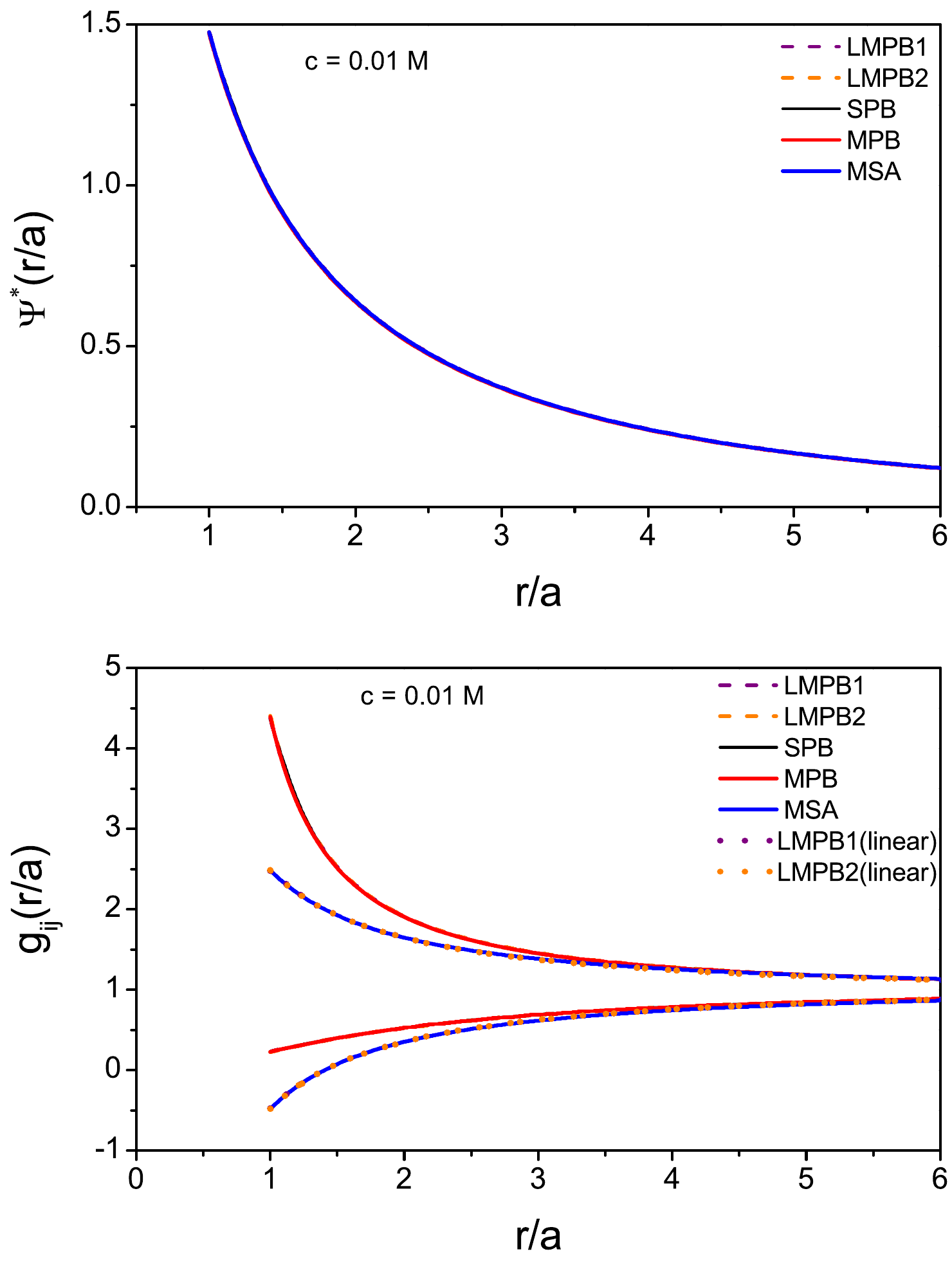}}
	\caption{(Colour online) The LMPB1, LMPB2, MSA, SPB, and MPB reduced mean
		electrostatic potential $\psi ^{*}(r/a) [ =(\beta |e|)\psi (r/a)]$ (upper panel) and the
		radial distribution function $g_{ij}(r/a)$ (lower panel) as functions of $r/a$ for a 1:1
		RPM electrolyte at $c =$~0.01~mol/dm$^{3})$ ($y =$~0.13985). The $g_{ij}(r/a)$ from the linearized LMPB1
		and LMPB2 theories are also shown in the lower panel.}
	\label{fig6}
\end{figure}

\begin{figure}[!t]
	\centerline{\includegraphics [height=4.5in]{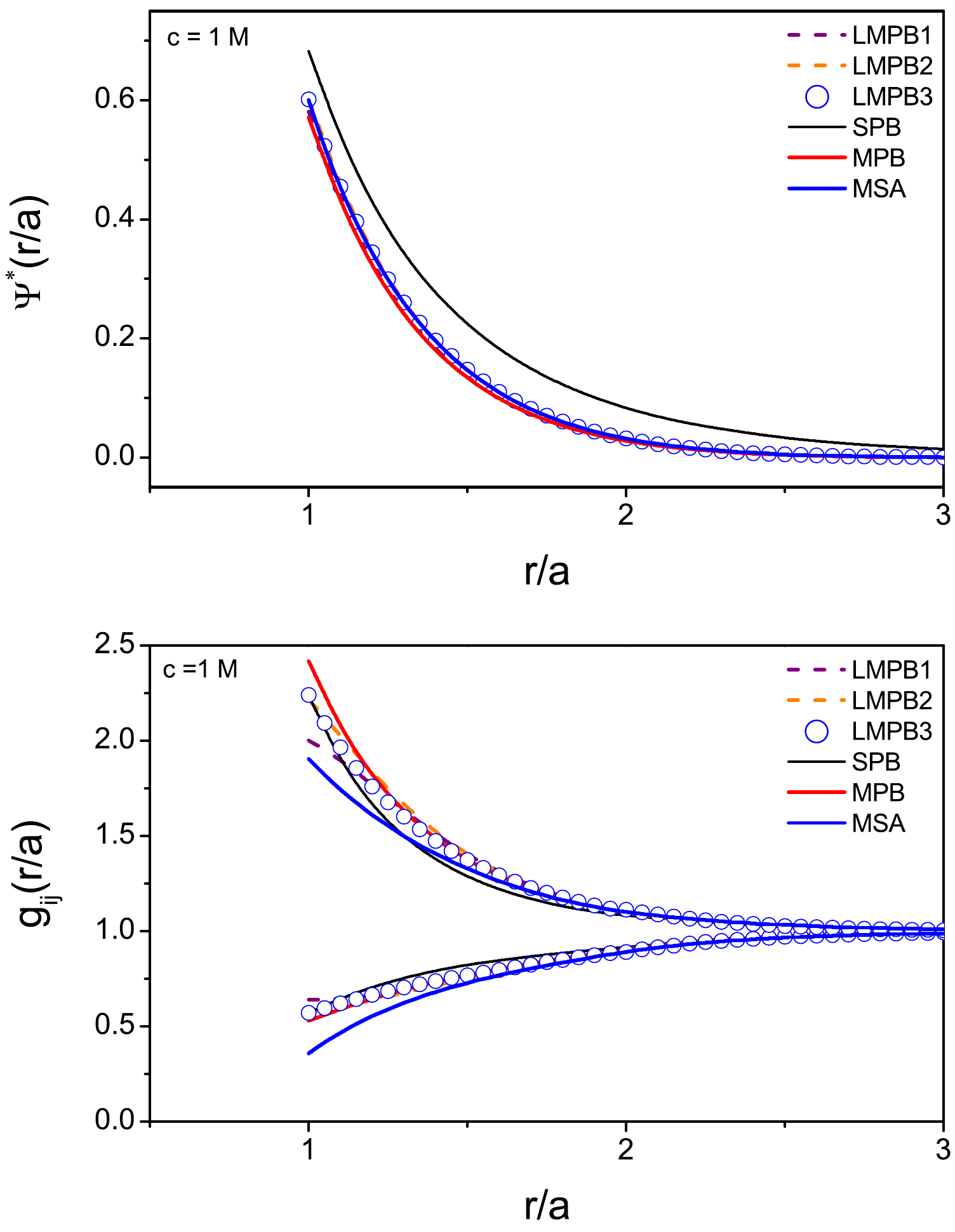}}
	\caption{(Colour online) The LMPB1, LMPB2, LMPB3, MSA, SPB, and MPB reduced mean
		electrostatic potential $\psi ^{*}(r/a) [ =(\beta |e|)\psi (r/a)]$ (upper panel) and the
		radial distribution function $g_{ij}(r/a)$ (lower panel) as functions of $r/a$ for a 1:1
		RPM electrolyte at $c =$~1~mol/dm$^{3}$ ($y =$~1.3985).}
	\label{fig7}
\end{figure}
    In the course of these calculations we have observed an equivalence
of the excess internal energy with the natural logarithm of the electrical
component of the mean activity coefficient for the LMPB and the MSA theories, viz.,

\begin{equation}
\frac{\beta U^{(\text{ex})}}{\rho } = \ln \gamma _{\pm }^{(\text{el})},
\label{m2_28}
\end{equation}

\noindent where $\rho = \sum _{s}\rho _{s}$.
    
    Here, we outline an argument why this is expected for a linear theory. Such an equivalence
has recently been observed by Kjellander with regard to his Multiple-Decay Extended Debye-H\"{u}ckel
MDE-DH theory of electrolytes~\cite{kjellander}.

    From equation~(\ref{m2_7}) we have

\begin{equation}
\psi _{i}(r) = \left(\frac{\rd u_{i}}{\rd r}\right)_{r=a} + \frac{|e|Z_{i}}{4\piup \varepsilon _{0} \varepsilon _{r}r}
\label{m2_29}
\end{equation}

\noindent with

\begin{equation}
\left(\frac{\rd u_{i}}{\rd r}\right)_{r=a} = a\left(\frac{\rd \psi _{i}}{\rd r}\right)_{r=a} + \frac{u_{i}(a)}{a} .
\label{m2_30}
\end{equation}

\noindent Using the Gauss's law for the electric field $-(\rd \psi _{i}/\rd r)_{r=a}$, and assuming that $u_{i}(a)$
is linear in $e_{i}(=|e|Z_{i})$, equation~(\ref{m2_29}) can be written as

\begin{equation}
\psi _{i}(r) = \frac{|e|Z_{i}}{4\piup \varepsilon _{0}\varepsilon _{r}r} +\frac{|e|Z_{i}C}{4\piup \varepsilon _{0}\varepsilon _{r}a},
\label{m2_31}
\end{equation}

\noindent where $C$ is independent of $e_{i}$. Combining now equations~(\ref{m2_22}) and~(\ref{m2_31}) we get

\begin{equation}
U^{(\text{ex})}  =  \frac{e^{2}C}{8\piup \varepsilon _{0}\varepsilon _{r}a}\sum _{s}Z_{s}^{2}\rho _{s} 
          =  -\frac{e^{2}\rho C Z_{+}Z_{-}}{8\piup \varepsilon _{0}\varepsilon _{r}a},
\label{m2_32}
\end{equation}

\noindent for a single electrolyte. In writing the above we have invoked the global neutrality $\sum _{s}Z_{s}\rho _{s} =$~0.
Now, from equations~(\ref{m2_24}) and~(\ref{m2_31}), we immediately have

\begin{equation}
\ln \gamma _{i}^{(\text{el})} = \frac {e^{2}\beta Z_{i}^{2}C}{8\piup \varepsilon _{0}\varepsilon _{r}a}.
\label{m2_33}
\end{equation}
Hence, using equation~(\ref{m2_27}) the mean activity can be written as

\begin{equation}
\ln \gamma _{\pm}^{(\text{el})} = -\frac{e^{2}\beta CZ_{+}Z_{-}}{8\piup \varepsilon _{0}\varepsilon _{r}a}.
\label{m2_34}
\end{equation}
 Equation~(\ref{m2_28}) now follows upon eliminating $C$ between the equations~(\ref{m2_32}) and~(\ref{m2_34}).

\section{Results and discussion}\label{s3}
\begin{figure}[!t]
	\centerline{\includegraphics [height=4.5in]{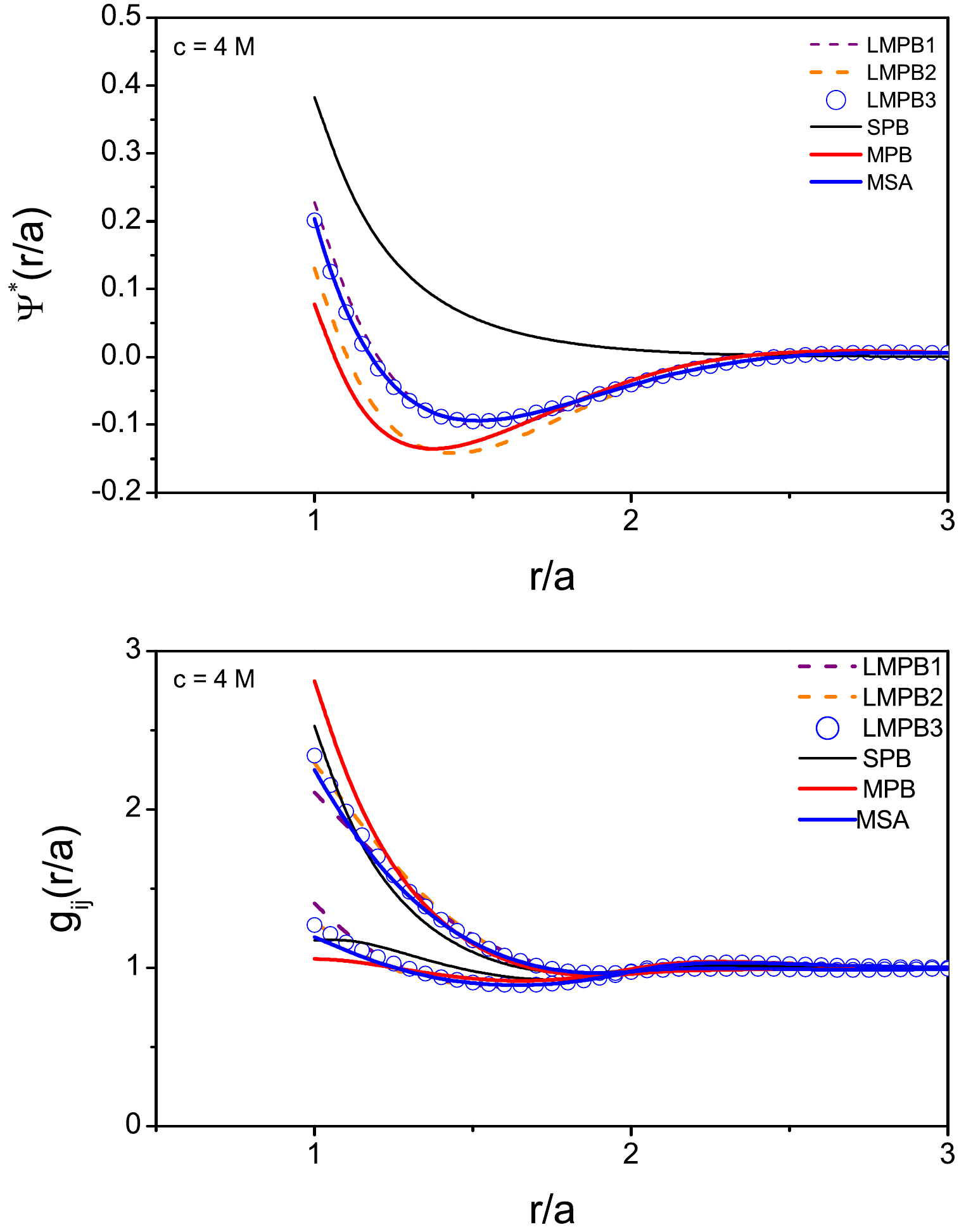}}
	\caption{(Colour online) The LMPB1, LMPB2, LMPB3, MSA, SPB, and MPB reduced mean
		electrostatic potential $\psi ^{*}(r/a) [ =(\beta |e|)\psi (r/a)]$ (upper panel) and the
		radial distribution function $g_{ij}(r/a)$ (lower panel) as functions of $r/a$ for a 1:1
		RPM electrolyte at $c =$~4~mol/dm$^{3}$ ($y =$~2.797).}
	\label{fig8}
\end{figure}

    In presenting the results we include data from the SPB, MPB,
and the MSA theories for comparison purposes. While the MSA results were obtained
from the relevant analytic expressions (see, for example, references~\cite{blum2,blum3,sanchezblum}),
the non-linear SPB and MPB equations were solved numerically using a quasi-linearization
technique~\cite{bellmankalaba} used successfully in earlier works~\cite{martinez,
molero,outh5,outh4}. The hard sphere part $g_{is}^{0}$ was approximated by the Percus-Yevick
uncharged pair distributions~\cite{leonardhendersonbarker,grundkehenderson} and
their corrections due to Verlet and Weis~\cite{verletweis}. In evaluating the
$g_{is}^{0}$ we used an efficient numerical technique developed by Perram~\cite{perram}.
Similarly, the hard sphere individual activity coefficient $\ln\gamma _{s}^{(\text{HS})}$
was determined using the formulations of Ebeling and Scherwinski~\cite{ebelsche}.

    Except for one case of asymmetric 2:1 valency system, all calculations reported
here are for 1:1 symmetric valency systems at temperature $T =$~298~K and relative
permittivity $\varepsilon =$~78.38, which corresponds to a water-like solvent. These
values are in line with that used in the MC simulations of Abbas et al.~\cite{abbas1,abbas2}.
The other physical parameters like the common ionic diameter $a$ and the concentration $c$
were variable and were fitted to the MC system being compared to.

\subsection{Thermodynamics}

    We begin this discussion by considering the results for $\phi $ and ln$\gamma _{\pm}$
in four 1:1 RPM salt solutions, viz., NaF, NaCl, HCl, and LiI shown in figures~\ref{fig1}, \ref{fig2}, \ref{fig3},
and \ref{fig4}, respectively. Abbas et al.~\cite{abbas1,abbas2} actually simulated over 100 PM and
RPM salts with different valencies 1:1, 2:1, and 3:1 and covering a wide range of ionic
sizes and solution concentrations. We have chosen these four since these encompass a fair
range of ionic size starting from $a =$~1.435~$\times~$10$^{-10}$~m for NaF (figure~\ref{fig1})
to $a =$~2.325~$\times~$10$^{-10}$~m for LiI (figure~\ref{fig4}). We have actually carried out
calculations for more 1:1 salts at their MC parameters, where the results show similar
characteristics and are hence not shown here for brevity. The four sets of results being
displayed also constitute a good representative sample.

\begin{figure}[!t]
	\centerline{\includegraphics [height=4.5in]{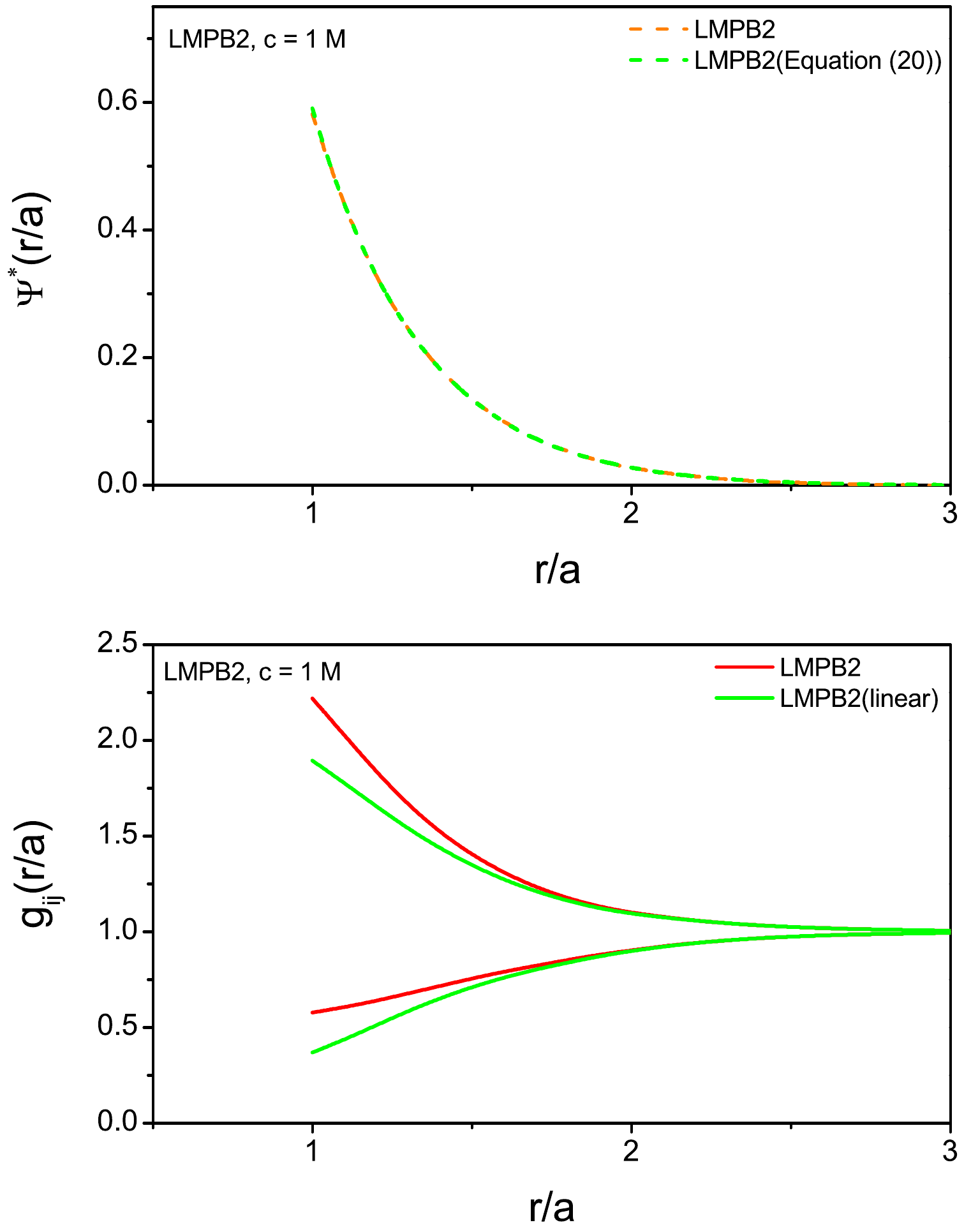}}
	\caption{(Colour online) The LMPB2 reduced mean electrostatic potential $\psi ^{*}(r/a) [ =(\beta |e|)\psi (r/a)]$
		together with the LMPB2 $\psi ^{*}(r/a)$ taken from equation~(\ref{m2_20}) (upper panel) and the LMPB2
		and LMPB2(linear) radial distribution function $g_{ij}(r/a)$ (lower panel)
		as functions of $r/a$ for a 1:1 RPM electrolyte at $c =$~1~mol/dm$^{3}$ ($y =$~1.3985).}
	\label{fig9}
\end{figure}

\begin{figure}[!t]
	\centerline{\includegraphics [height=4.5in]{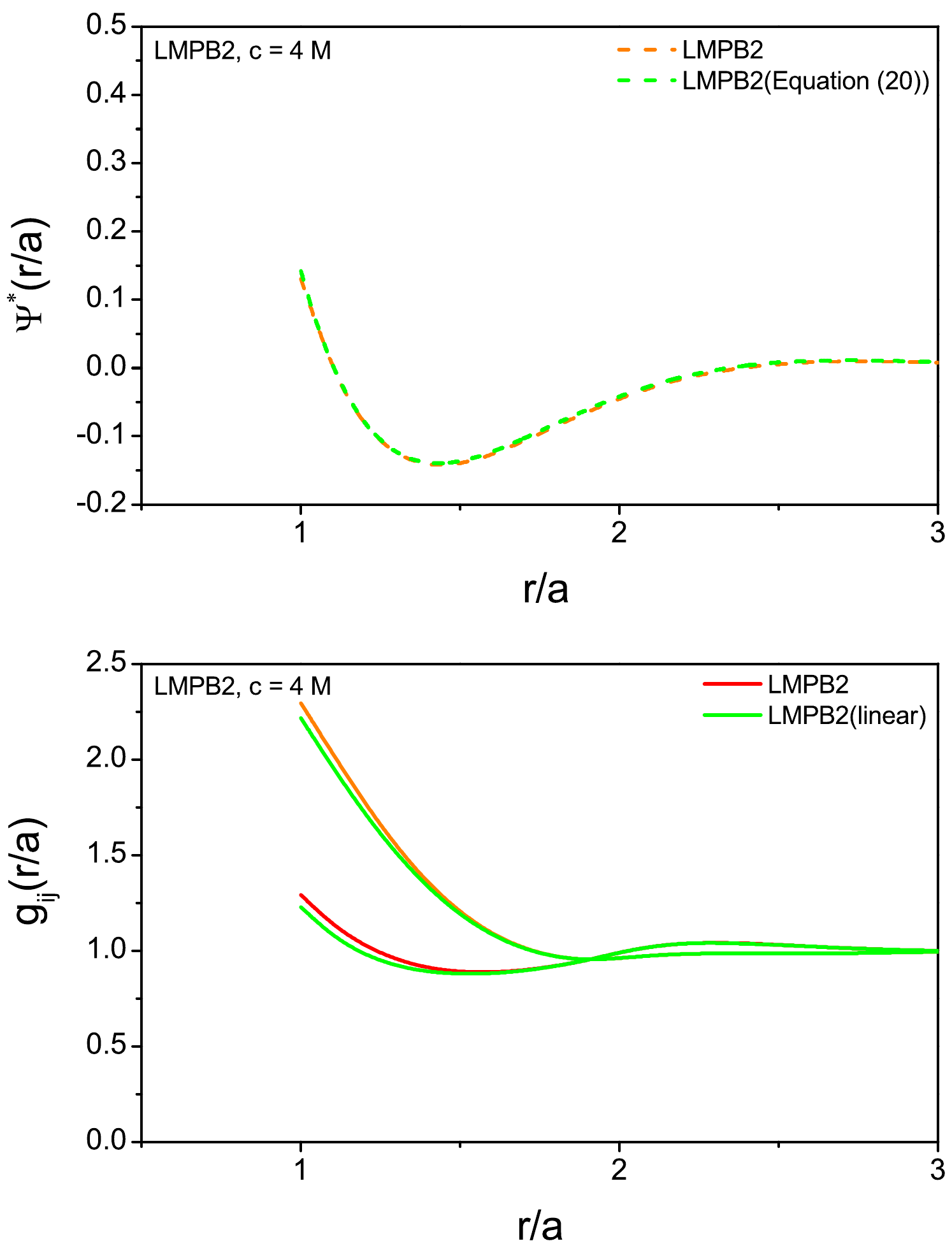}}
	\caption{(Colour online) The LMPB2 reduced mean electrostatic potential $\psi ^{*}(r/a) [ =(\beta |e|)\psi (r/a)]$
		together with the LMPB2 $\psi ^{*}(r/a)$ taken from equation~(\ref{m2_20}) (upper panel) and the LMPB2
		and LMPB2(linear) radial distribution function $g_{ij}(r/a)$ (lower panel)
		as functions of $r/a$ for a 1:1 RPM electrolyte at $c =$~4~mol/dm$^{3}$ ($y =$~2.797).}
	\label{fig10}
\end{figure}

    A striking feature in figures~\ref{fig1}--\ref{fig4} is the remarkable consistency of the LMPB curves
both among themselves and with the MC data. Indeed, the results of all the theories including
that from the MSA and the MPB are in very close agreement with each other with only the SPB
ln$\gamma _{\pm}$ showing some deviation at the two lower diameters (figures~\ref{fig1} and \ref{fig2}).
This behaviour pattern carries over to the 2:1 valency (MgCl$_{2}$) situation in figure~\ref{fig5}
where again the LMPB theories are at par with the formal theories in reproducing the MC data.

\subsection{Structure}

    The structural results are presented in figures~\ref{fig6}--\ref{fig10}. For these calculations, we have
used a fixed value of $a =$~4.25~$\times $10$^{-10}$~m. Figure~\ref{fig6} shows the LMPB1, LMPB2,
SPB, MPB, and MSA results for the reduced mean electrostatic potential $\psi ^{*} (= |e|\beta \psi = |e|\beta)u/r)$
(upper panel) and the $g_{ij}$ (lower panel) at the electrolyte concentration $c =$~0.01~mol/dm$^{3}$
($y =$~0.13985). For the $\psi ^{*}$, the various curves are indistinguishable from each other,
which is also the case with the $g_{ij}$'s except for the MSA curves. This occurs owing
to the linear nature of the MSA. The contact values of the MSA $g_{ij}$'s are underestimated
and generally  the MSA curves lie below the others. The unphysical negative values of the
MSA coion $g$ around the contact are noticeable. This is of course a known shortcoming of the
MSA at low  concentrations. In the lower panel, we have also plotted the corresponding LMPB1(linear)
and LMPB2(linear) $g_{ij}$'s. It is interesting, although perhaps not surprising, to note that
these linear $g_{ij}$'s follow the MSA very closely leading to the coion $g$'s also
becoming negative at and near the contact.

    The results at the higher concentrations of $c =$~1~mol/dm$^{3}$ ($y =$~1.3985) and
$c =$~4~mol/dm$^{3}$ ($y =$~2.797) are displayed in figures~\ref{fig7} and \ref{fig8}, respectively.
In both situations $y > y_{c}$, and we see the beginnings of oscillations in the LMPB,
MPB, and MSA profiles in figure~\ref{fig7}, which, expectedly, become more pronounced in figure~\ref{fig8}.
Although not quite as quantitative as they are at the lower $c =$~0.01~mol/dm$^{3}$ in
figure~\ref{fig6}, the LMPB predictions continue to show good qualitative agreement overall with
that from  the MPB at these enhanced concentrations. Nonetheless, the differences in the
structure can produce discrepancies in various properties such as thermodynamics via different
routes and transport properties, especially at higher concentrations and valencies.

The oscillations in the LMPB curves
are due to the fact that linearization of the MPB equation retains the aspects of the
fluctuation  potential terms. By contrast, no such terms  occur in the classical SPB and
hence no oscillations.

    A consistency check on the LMPB $g_{ij}$'s and a comparison of these $g_{ij}$'s with
their linear version are shown in figures~\ref{fig9} and \ref{fig10} at $c =$~1~mol/dm$^{3}$ ($y =$~1.3985)
and $c =$~4~mol/dm$^{3}$ ($y =$~1.3985), respectively. For illustrative purposes we only
show the results for LMPB2 since the other LMPB1 and LMPB3 results are very similar. In the upper
panels, the LMPB2 $\psi ^{*}$ from equations~(\ref{m2_14}) and (\ref{m2_20}) are plotted. The consistency
of these $\psi {*}$ at both of these concentrations is noteworthy, and points, in turn, to
the consistency of the approximation made in equation~(\ref{m2_18}). A similar effect is observed with
regard to the non-linear LMPB2 and the linear LMPB2(linear) $g_{ij}$'s, viz., equations~(\ref{m2_18}) and (\ref{m2_19}),
in the lower panels with the linear $g_{ij}$'s showing some discrepancy only near the contact.

\section{Conclusion}\label{s4}

   This work represents a continuation of our earlier study, \cite{outh10}, on the applicability
of linear modified Poisson-Boltzmann theory to the electrolyte solution theory.
The main achievement of this paper is the characterization of thermodynamics
of RPM electrolytes using the set of three linear modified Poisson-Boltzmann
theories proposed in~\cite{outh10}. The osmotic and mean activity coefficients predicted by the
LMPB1, LMPB2, and LMPB3 theories for model electrolytes mimicking NaF, NaCl, HCl, LiI,
and MgCl$_{2}$ solutions are consistent among themselves, and show a very good agreement
with those from the  MPB and the MSA theories. The linear results also reproduce the MC
data for these systems to a comparable degree of accuracy.

    We have also studied structural aspects of RPM electrolytes at different concentrations
through the mean electrostatic potential and the radial distribution functions as revealed by
the LMPB theories. Again, the results show an overall qualitative or better level of correspondence
with the MPB and MSA data. A notable feature of the LMPB $g_{ij}$ and $\psi _{i}^{*}$ curves
is that they show oscillations at higher solution concentrations. These oscillations are manifestations
of inter-ionic correlations and occur since linearization of the MPB equation retains the aspects of the
(MPB) fluctuation terms in the LMPB equations as seen in~\cite{outh10}. In contrast, the classical mean-field
theories such as the DH and SPB do not incorporate ionic correlations and as such fail to capture
such oscillations.

    An interesting finding of the present work is the equivalence of the excess internal energy and
the electrical contribution to the mean activity coefficient for linear theories. In the course of our
calculations we have found this to be true of the LMPB1, LMPB2, LMPB3, and the MSA --- all linear
theories, and under all physical conditions. This has also been observed by Kjellander~\cite{kjellander}
with his MDE-DH theory of electrolytes. We have outlined here a general argument as to why
this should necessarily be so for all linear theories. This exact result is likely to be useful when
working with such theories. For the non-linear SPB and MPB cases, however, we have found the numerical
values of these two quantities to be close, but not identical.

    The results of this study give a practical relevance to the LMPB approach. In some sense, the LMPB
theories are intermediate between the DH/DHLL theories and the MPB or other formal statistical mechanical
theories. There are parallels to the DH/DHLL in that many of the thermodynamic and structural quantities
of interest in the LMPB theories are analytical. However, unlike the former theories, the latter incorporate
the ionic exclusion volume and correlation terms and hence improve the accuracy of their
results relative to the DH/DHLL. These features of the LMPB theories can be useful in the routine, everyday
analysis of experimental data. Similarly, in the analysis of numerically intensive theoretical problems
(see for example, reference~\cite{ulloa}) the use of LMPB analytical expressions as initial input in
iterative processes can prove to be useful. Furthermore, the success of the LMPB approach in analyzing
the thermodynamics of 1:1 valency systems makes this a potentially attractive method that can be used to
explore more complex situations such as higher valency systems~\cite{kjellander} and/or systems with a
variable dielectric constant~\cite {abbas3}. Studies of higher valency and mixed electrolyte
systems would also be useful in order to see the limitations of the LMPB theories.

\section*{Acknowledgements}

    We are grateful to Professor C.W. Outhwaite of the University of Sheffield, UK, for his support
during the course of this work, numerical checks of some of the calculations, a critical reading of the manuscript,
and many helpful comments.

\ukrainianpart

\title
{Структурні та термодинамічні властивості систем у рамках лінійної модифікованої теорії Пуассона-Больцмана для примітивної моделі електролітів}
\author[Л. Б. Буіян]{Л. Б. Буіян}
\address{Лабораторія теоретичної фізики, Фізичний факультет університету Пуерто-Ріко, 17 Авеніда Універсидад, STE 1701, Сан Хуан, Пуерто-Ріко 00925-2537, США
}

\makeukrtitle

\begin{abstract}
Структура та термодинаміка систем в рамках обмеженої примітивної моделі
електроліту досліджуються за допомогою трьох нещодавно розроблених варіантів лінійної форми модифікованого рівняння Пуассона-Больцмана. Отримані аналітичні вирази для осмотичного коефіцієнта та електричної складової середнього коефіцієнта активності. Результати для осмотичного коефіцієнта та середнього значення активності порівнюються з даними для інших різновидів середньосферичного наближення, симетричної та модифікованої теорій Пуассона-Больцмана, а також з наявними результатами моделювання Монте-Карло. Лінійні теорії надзвичайно точно передбачають термодинаміку систем у порівнянні з комп'ютерним моделюванням і узгоджуються із середньосферичним наближенням та модифікованими результатами Пуассона-Больцмана. Прогнозована структура, представлена у вигляді радіальних функцій розподілу та середнього електростатичного потенціалу, також добре узгоджується з відповідними результатами суміжних теорій. Показано, що надлишкова внутрішня енергія та електрична складова середнього коефіцієнта активності аналітично ідентичні для середньосферичного наближення та лінійно модифікованих теорій Пуассона-Больцмана.
	\keywords обмежена примітивна модель, структура, осмотичний коефіцієнт, коефіцієнт активності,
	лінійна модифікована теорія Пуассона-Больцмана
\end{abstract}


\begin{thebibliography}{10}
\bibitem{debyehuckel}  Debye P., H\"{u}ckel E., {Z. Phys.}, 1923, {\bf 24}, 185.
\bibitem{mcquarrie} McQuarrie D.~A., {Statistical Mechanics}, Harper and Row, New York, 1976.
\bibitem{kirkwood} Kirkwood J.~G., J. Chem. Phys., 1934, {\bf 2}, 767, \doi{10.1063/1.1749393}.
\bibitem{card} Card D.~N.,   Valleau J.~P., J. Chem. Phys., 1970, {\bf 52}, 6232, \doi{/10.1063/1.1672932}.
\bibitem{rasaiah} Rasaiah J.~C., Card D.~N.,  Valleau, J.~P., J. Chem. Phys., 1972, {\bf 56}, 248,
    \doi{10.1063/1.1676854}.
\bibitem{valleau1} Valleau J.~P., Cohen L.~K., J. Chem. Phys., 1980, {\bf 72}, 5935, \doi{10.1063/1.439092}.
\bibitem{valleau2} Valleau J.~P., Cohen L.~K., Card D.~N., J. Chem. Phys., 1980, {\bf 72}, 5942, \doi{10.1063/1.439093}.
\bibitem{rogde} Rogde S.~A., Chem. Phys. Lett., 1983, {\bf 103}, 133, \doi{10.1016/0009-2614(83)87480-6}.
\bibitem{abramo} Abramo M.~C.,  Caccamo C., Malescio G., Pizzimenti G., Rogde S.~A., J. Chem. Phys., 1984,
    {\bf 80}, 4396,\\ \doi{10.1063/1.447217}.
\bibitem{vlachy} Vlachy V., Annu. Rev. Phys. Chem., 1999, {\bf 50}, 145, \doi{10.1146/annurev.physchem.50.1.145}.
\bibitem{levin} Levin Y., Rep. Prog. Phys., 2002, {\bf 65}, 1577, \doi{10.1088/0034-4885/65/11/201}.
\bibitem{henderson} Henderson D.,  Holovko M., Trokhymchuk A., (Eds.), Ionic Soft Matter: Modern Trends in Theory and Applications, NATO Science Series II: Mathematics, Physics and Chemistry Series, Vol.~206, Springer, Dordrecht, 2004, \doi{10.1007/1-4020-3659-0}.
\bibitem{messina} Messina R., J. Phys.: Condens. Matter, 2009, {\bf 21}, 113102, \doi{10.1088/0953-8984/21/11/113102}.
\bibitem{kalyuzhnyi}  Kalyuzhnyi V., Vlachy V., Dill K.~A.,  Phys. Chem. Chem. Phys., 2010, {\bf 12}, 6260, \doi{10.1039/b924735a}.
\bibitem{chersty} Chersty A.~G., Phys. Chem. Chem. Phys., 2011, {\bf 13}, 9942, \doi{10.1039/c0cp02796k}.
\bibitem{outh1} Outhwaite C.~W., J. Chem. Phys., 1969, {\bf 50}, 2277, \doi{10.1063/1.1671378}.
\bibitem{outh2} Outhwaite C.~W., Mol. Phys., 1974, {\bf 28}, 217, \doi{10.1080/00268977400101651}.
\bibitem{outh8} Outhwaite C.~W., In: Statistical Mechanics (Specialist Periodical Report), Singer K. (Ed.),
    The Chemical Society, London, 1975, vol. 2, ch. 3, p. 188--255.


\bibitem{martinez} Martinez M.~M., Bhuiyan L.~B., Outhwaite C.~W., J. Chem. Soc., Faraday Trans., 1990.
    {\bf 86}, 3383, \\\doi{10.1039/FT9908603383}.

\bibitem{molero} Molero M., Outhwaite C.~W.  Bhuiyan L.~B., J. Chem. Soc., Faraday Trans., 1992, {\bf 88}, 1541, \\\doi{10.1039/FT9928801541}.
\bibitem{outh5} Outhwaite C.~W., Molero M., Bhuiyan L.~B., J. Chem. Soc., Faraday Trans., 1993, {\bf 89}, 1315,
   \\ \doi{10.1039/FT9938901315}.
\bibitem{outh6} Outhwaite C.~W., Condens. Matter Phys., 2004, {\bf 7}, 719, \doi{10.5488/CMP.7.4.719}.
\bibitem{friedman1} Friedman H.~L., Ionic Solution Theory, Wiley, New York, 1962.
\bibitem{friedman2}Friedman H.~L., A Course in Statistical Mechanics, Prentice-Hall, New Jersey, 1985.
\bibitem{blum1} Blum L., Mol. Phys., 1975, {\bf 30}, 1529, \doi{10.1080/00268977500103051}.
\bibitem{outh7} Outhwaite C.~W., Hutson V.~C.~L., Mol. Phys., 1975, {\bf 29}, 1521, \doi{10.1080/00268977500101331}.
\bibitem{blum2} Blum L., Theoretical Chemistry, Advances and Perspectives, Vol.5,  Eyring H., Henderson H.,
(Eds.), Academic Press, New York, 1980.
\bibitem{attard} Attard P., Adv. Chem. Phys., 1996, {\bf 92}, 1, \doi{10.1002/9780470141519.ch1}.
\bibitem{hansenlowen} Hansen J.-P., L\"{o}wen H., Annu. Rev. Phys. Chem., 2000, {\bf 51}, 209, \doi{10.1146/annurev.physchem.51.1.209}.
\bibitem{outh9} Outhwaite C.~W., Mol. Phys., 1971, {\bf 20}, 705, \doi{10.1080/00268977100100671}.
\bibitem{outh10} Outhwaite C.~W., Bhuiyan L.~B., Condens. Matter Phys., 2019, {\bf 22}, 23801 (14 pages), \\\doi{10.5488/CMP.22.23801}.
\bibitem{outh4} Outhwaite C.~W., Molero M., Bhuiyan L.~B., J. Chem. Soc., Faraday Trans., 1991, {\bf 87}, 3227,
   \\ \doi{10.1039/FT9918703227}.
\bibitem{barthelkrienkekunz} Barthel J.~M.~G.,  Krienke H.,  Kunz W.,~In: Topics in Physicsl Chemistry 5, Baumg\"{a}rte H.,
Franck, E. U., Gr\"{u}nbein~E., (Eds.), Springer, New York, 1998.
\bibitem{collinsneilsonenderby} Collins K.~D., Neilson G.~W., Enderby J.~E., Biophys. Chem., 2007, {\bf 128}, 95, \doi{10.1016/j.bpc.2007.03.009}.
\bibitem{quinones} Qui\~{n}ones A.~O., Bhuiyan L.~B., Outhwaite C.~W., Condens. Matter Phys., 2018, {\bf 21}, 23802 (1--10 pages), \doi{10.5488/CMP.21.23802}.
\bibitem{abbas1}  Abbas Z.,  Ahlberg E.,  Nordholm S., Fluid Phase Equilib., 2007, {\bf 260}, 233, \doi{10.1016/j.fluid.2007.07.026}.
\bibitem{abbas2} Abbas Z., Ahlberg E., Nordholm S., J. Phys. Chem. B, 2009, {\bf 113}, 5905, \doi{10.1021/jp808427f}.
\bibitem{robinson} Robinson R. A,  Stokes R.~H., Electrolyte Solutions, Second Revised Edition, Dover, New York, 2002.
\bibitem{stillinger} Stillinger F.~H., Lovett R., J. Chem. Phys., 1968, {\bf 48}, 3858, \doi{10.1063/1.1669709}.
\bibitem{outh11} Outhwaite C.~W., Chem. Phys. Lett., 1974, {\bf 24}, 73, \doi{10.1016/0009-2614(74)80216-2}.
\bibitem{carley} Carley D.~D., J. Chem. Phys., 1967, {\bf 46}, 3783, \doi{10.1063/1.1840451}.
\bibitem{outh12} Outhwaite C.~W., Chem. Phys. Lett., 1976, {\bf 37}, 383, \doi{10.1016/0009-2614(76)80238-2}.
\bibitem{ebelsche} Ebeling W., Scherwinski K., Z. Phys. Chem. (Leipzig), 1983, {\bf 264}, 1, \doi{10.1515/zpch-1983-26402}.
\bibitem{kjellander} Kjellander R., Phys. Chem. Chem. Phys., 2020, {\bf 22}, 23952, \doi{10.1039/d0cp02742a}.
\bibitem{blum3} Blum L., Hoye J.~S., J. Phys. Chem., 1977, {\bf 30}, 1529, \doi{10.1021/j100528a019}.
\bibitem{sanchezblum} Sanchez-Castro C., Blum L., J. Phys. Chem., 1989, {\bf 93}, 7478, \doi{10.1021/j100358a043}.

\bibitem{bellmankalaba} Bellman R., Kalaba R., Quasilinearization  Nonlinear Boundary Value Problems,
    Elsevier, New York, 1965.
\bibitem{leonardhendersonbarker} Leonard P.~J., Henderson D., Barker J.~A., Mol. Phys., 1971, {\bf 21}, 107, \doi{10.1080/00268977100101221}.
\bibitem{grundkehenderson} Grundke E.~W., Henderson D., Mol. Phys., 1974, {\bf 24}, 269, \doi{10.1080/00268977200101431}.
\bibitem{verletweis} Verlet L., Weis J.~J., Phys. Rev. A, 1972, {\bf 5}, 939, \doi{10.1103/physreva.5.939}.
\bibitem{perram} Perram J.~W., Mol. Phys., 1975, {\bf 30}, 1505, \doi{10.1080/00268977500103021}.

\bibitem{ulloa} Ulloa-D\'{a}vila E.~O., Bhuiyan L.~B., Condens. Matter Phys., 2017, {\bf 20}, 43801\\ (16 pages), \doi{10.5488/CMP.20.43801}.
\bibitem{abbas3} Abbas Z., Ahlberg E., J. Sol. Chem., 2019, {\bf 48}, 1222, \doi{10.1007/s10953-019-00905-y}.
\end{thebibliography}
\end{document}